\documentclass[preprint,prl,aps]{revtex4-1}
\usepackage{graphicx}

\begin{document}
\title{An Interpretation of Quantum Foundations Based on Density Functional Theory and Polymer Self-Consistent Field Theory}
\author{Russell B. Thompson}
\email{thompson@uwaterloo.ca}
\affiliation{Department of Physics \& Astronomy and Waterloo Institute for Nanotechnology, University of Waterloo, 200 University Avenue West, Waterloo, Ontario, Canada N2L 3G1}
\date{26 April 2021}

\begin{abstract}
The Feynman quantum-classical isomorphism between classical statistical mechanics in 3+1 dimensions and quantum statistical mechanics in 3 dimensions is used to connect classical polymer self-consistent field theory with quantum time-dependent density functional theory. This allows the theorems of density functional theory to relate non-relativistic quantum mechanics back to a classical statistical mechanical derivation of polymer self-consistent field theory for ring polymers in a 4 dimensional thermal-space. One dynamic postulate is added to two static postulates which allows for a complete description of quantum physics from a 5 dimensional thermal-space-time ensemble perspective which also removes the measurement problem. In the classical limit, a cylinder condition naturally arises as the thermal dimension becomes irrelevant, providing a justification for using 5 dimensions and a cylinder condition in general relativity, which is known to produce 4 dimensional space-time gravity and Maxwell's equations. Thus, in this approach, the postulates of electromagnetism become derived results of a special case of a ring polymer interpretation of quantum foundations.
\end{abstract}

\maketitle

It has been known for almost seventy years that equilibrium properties of many body quantum systems can be calculated using classical statistical mechanics through an isomorphism between quantum theory and ring polymers. The mathematics of this isomorphism was presented by Feynman in 1953 \cite{Feynman1953a,Feynman1953b,Feynman1953c}, an explicit identification with ring polymers was made by Chandler and Wolynes in 1981 \cite{Chandler1981}, and this ring polymer isomorphism now forms the basis of path integral simulations \cite{Ceperley1995,Habershon2013,Voth1996,Roy1999a,Roy1999b}. The idea is that in quantum statistical mechanics, the partition function for quantum particles is expressed in terms of path integrals weighted by thermal ``trajectories'' running from zero to $\beta = 1/k_BT$, where $k_B$ is Boltzmann's constant and $T$ is the temperature. Feynman derived his exact partition function by treating $\beta$ as an imaginary time variable \cite{Feynman1953b}, and although he didn't identify the thermal trajectory with classical ring polymers at the time, this has since become standard practice \cite{Chandler1981,Ceperley1995}. In this context, ``polymer'', does not refer literally to a macromolecule, but rather to an extended, non-point-like contour. Actual polymers can be parametrized by a space-curve ${\bf r}(s)$ which gives the location in space ${\bf r}$ for a position $s$ along the contour of the polymer. Quantum particles are parametrized by a thermal-curve ${\bf r}(s)$ which gives the expected location ${\bf r}$ in space at a temperature $s=\beta$. For quantum systems, one includes only trajectories for which ${\bf r}(0) = {\bf r}(\beta)$, so this maps onto ring polymers.

Another powerful tool for studying quantum many body systems is density functional theory (DFT). In previous work \cite{Thompson2019,Thompson2020}, it was shown that a theoretical polymer physics tool from equilibrium statistical mechanics called self-consistent field theory (SCFT), when applied to a system of ring polymers, gives identical mathematics as a quantum system in DFT. In other words, classical SCFT for ring polymers is isomorphic with quantum DFT. This is not surprising given the known isomorphism between classical and quantum partition functions just reviewed. An advantage of the SCFT perspective is that it is derived from first principles without using the theorems of DFT. These theorems basically state that there is a one-to-one mapping between the DFT representation of quantum systems in terms of one-body density functions and the underlying wave function representation \cite{Hohenberg1964,Mermin1965,Jones2015}. Based on the mathematics of the first principles SCFT derivation, one can instead postulate directly that quantum particles are thermal contours in a classical four dimensional thermal-space, with the extra dimension being the inverse thermal energy $\beta$. The DFT theorems then guarantee that all predictions will be consistent with known quantum mechanics. One is thus replacing the usual set of wave function-based postulates with a smaller number of assumptions in the 4D classical statistical mechanics interpretation. 

It's a difficult thing to measure the explanatory power of a theory based on counting postulates; there are normally about a half dozen postulates in non-relativistic quantum mechanics, but this number can be reduced depending on the presentation. For static, non-relativistic quantum mechanics, it was already shown that only two postulates are needed in the SCFT ring polymer derivation \cite{Thompson2019,Thompson2020}. The first is that quantum particles are fractal Gaussian threads in four dimensions (for the time independent case). This is equivalent to postulating the Heisenberg uncertainty relation in three dimensions \cite{Thompson2020}. The second postulate is that, for more than two threads, the contours have classical excluded volume in static 4D thermal-space. This assumption was shown to produce the correct shell structure for atoms expected from the Pauli exclusion principle, and so is a 4D replacement for the exclusion principle in 3D \cite{Thompson2020}. Most standard introductions to quantum mechanics don't include the exclusion principle as an explicit postulate, but Kaplan \cite{Kaplan2013} has demonstrated that it cannot be proven through ideas of indistinguishability as is often suggested. Thus, for the static case, the number of postulates is so significantly reduced by adopting a ring polymer picture of quantum particles that this is evidence for suggesting it has more explanatory power than other quantum interpretations. Additionally, since the ring polymer model arises within classical statistical mechanics (both postulates are entirely classical), it is necessarily an ensemble (statistical) interpretation, so there is no measurement problem (no collapse of a wavefunction). This feature of the ensemble interpretation has been exhaustively examined by Ballentine \cite{Ballentine1970,Ballentine2003,Ballentine2019} and has wide support -- see for example Bransden and Joachain \cite{Bransden2000}, Aharanov \cite{Aharanov2017}, Popper \cite{Popper1967}, Blokhintsev \cite{Blokhintsev1968} and Einstein \cite{Einstein1949}. The ring polymer model is more specific than the generic ensemble interpretation however, since ``hidden variables'' are clearly identified as thermal correlations, that is, the degrees of freedom of the nonlocal polymer in the higher dimensional space \cite{Thompson2020}.

The aim of this paper is to show that the classical 4D thermal-space ring polymer model of static quantum mechanics is consistent with 5D thermal-space-time quantum physics. That is, all predictions of time-dependent quantum mechanics can be shown to arise from a classical model of contours in a 5D thermal-space-time. This will be shown by introducing one more postulate, for a total of three, and applying the Runge-Gross theorem of time-dependent density functional theory (TDDFT). It then immediately follows that the postulates of electromagnetism, namely Maxwell's equations, arise spontaneously from these three postulates, reducing the number of foundational assumptions within physics more broadly. This connection with electromagnetism is achieved through the general relativistic ideas of Kaluza \cite{Kaluza1921}, since the thermal-space-time approach physically identifies a fifth dimension and has the ``cylinder condition'' of Kaluza built-in for the classical limit, as will be explained.

It is helpful to begin by briefly reviewing the static case. The classical statistical mechanics derivation of SCFT equations for general polymers can be found in references \cite{Matsen2020,Matsen2006,Qiu2006,Fredrickson2002,Fredrickson2006,Schmid1998} and for ring polymers specifically in reference \cite{Kim2012}. One can also obtain the same equations through quantum statistical mechanics as shown in reference \cite{Thompson2019}. These derivations will not be repeated here, but the main results will be summarized. 

From equilibrium statistical mechanics in the canonical ensemble with $N$ quantum particles (polymer-like Gaussian threads) in a volume $V$ at a temperature $T$, the free energy will be \cite{Thompson2019,Thompson2020}
\begin{equation}
F[n,w] = -\frac{N}{\beta}\ln Q(\beta) - \int d{\bf r} w({\bf r},\beta) n({\bf r},\beta) + U[n]    \label{FE1}
\end{equation}
which is a functional of the quantum particle density $n({\bf r},\beta)$ and a field $w({\bf r},\beta)$. $U[n]$ is a potential energy functional of the density and $Q(\beta)$ is a single particle partition function, defined below. Expressions for both the density and the field are obtained by setting functional derivatives of (\ref{FE1}) with respect to $n({\bf r},\beta)$ and $w({\bf r},\beta)$ equal to zero, giving
\begin{eqnarray}
n({\bf r},\beta) &=& \frac{n_0}{Q(\beta)} q({\bf r},{\bf r},\beta)    \label{n3}  \\
w({\bf r},\beta) &=& \frac{\delta U[n]}{\delta n({\bf r},\beta)}  \label{w3}
\end{eqnarray}
where $n_0 = N/V$ is the overall density. The single particle partition function is given by
\begin{equation}
Q(\beta) = \frac{1}{V} \int d{\bf r} q({\bf r},{\bf r},\beta)   \label{Q3}
\end{equation}
where a propagator $q({\bf r}_0,{\bf r},\beta)$ is a solution to the modified diffusion equation
\begin{equation}
\frac{\partial q({\bf r}_0,{\bf r},\beta)}{\partial \beta} = \mathcal{H} q({\bf r}_0,{\bf r},\beta)  \label{diff3}
\end{equation}
with
\begin{equation}
\mathcal{H} =  \frac{\hbar^2}{2m} \nabla^2 - w({\bf r},\beta)   \label{H1}
\end{equation}
subject to the ``initial'' conditions
\begin{equation}
q({\bf r}_0,{\bf r},0) = V \delta({\bf r}-{\bf r}_0)  .  \label{init3}
\end{equation}
This set of equations can be solved when the potential $U[n]$, which describes the system, is given. For example, for an atomic system, $U[n]$ would include Coulomb terms for the ionic nucleus and electron-electron interactions, as well as electron self-interaction corrections and a Pauli term to enforce the exclusion principle. 

Appendix B of reference \cite{Thompson2019} shows that this set of SCFT equations reduces to Kohn-Sham DFT \cite{Kohn1965} assuming a perfect enforcement of the exclusion principle. The eigenvalues $\varepsilon_i$ and eigenfunctions $ \phi_i({\bf r})$ of the operator $\mathcal{H}$ of equation (\ref{H1}) are given by
\begin{equation}
\mathcal{H} \phi_i({\bf r}) = \varepsilon_i \phi_i({\bf r})      .   \label{KS1}
\end{equation}
The modified diffusion equation (\ref{diff3}) is expanded in terms of the eigenfunctions of $\mathcal{H}$ so that the density (\ref{n3}) becomes \cite{Thompson2019,Matsen2006}
\begin{equation}
n({\bf r}) = \frac{1}{V}\sum_{i=1}^\infty f(\varepsilon_i - \mu) \left| \phi_i ({\bf r})\right|^2  \label{n1}
\end{equation}
where $f(\varepsilon_i - \mu)$ is the Fermi-Dirac distribution and $\mu$ is the chemical potential. In the zero temperature limit, this becomes the familiar Kohn-Sham expression for the density.\footnote{The extra factor of $1/V$ is due to following the orthonormality definition of Matsen \cite{Matsen2006}.} Together with the Kohn-Sham equation (\ref{KS1}) and an expression for the field $w({\bf r},\beta)$ from equation (\ref{w3}), the DFT equations are complete. Thus, through the theorems of DFT \cite{Hohenberg1964,Mermin1965,Jones2015}, the two postulates of the SCFT formalism are formally able to reproduce all predictions of static, non-relativistic quantum physics from classical statistical mechanics in 4D.

Quantum mechanics is, however, fundamentally a dynamic theory. To incorporate dynamics into the SCFT methodology, one could use a direct generalization of the static derivation, although this ``frontal assault'' could be very difficult. It would involve a first principles non-equilibrium derivation of a dynamic action instead of a free energy. Then, one would vary the action to get a set of non-equilibrium equations describing quantum dynamics. One possible route for this could involve the Keldysh formalism \cite{vanLeeuwen2006b}. Other possibilities from polymer physics are the dynamic partition function approach used by Grzetic, Wickham and Shi \cite{Grzetic2014,Grzetic2020} and a similar method by Fredrickson and Orland \cite{Fredrickson2014}. Such methods still require postulating the dynamical behaviour of the thermal threads as input for the dynamics and inevitably require many approximations. This is a familiar situation in polymer physics where dynamic generalizations of  SCFT compromise conservation laws and often require working close to equilibrium --- see for example references \cite{Grzetic2014,Grzetic2020,Fredrickson2014,Fraaije1999,Hall2006,Hall2007,Muller2005,Qiu2008}. Whatever the method used, the results would need to be shown to reduce to a form of TDDFT so that the Runge-Gross theorem, which shows that a dynamic one-particle density can be used as the fundamental variable in quantum mechanics, would guarantee equivalent physical predictions to the wave function approach. This is the time-dependent version of the process used in the static case \cite{Thompson2019,Thompson2020}. The frontal assault is only necessary however if one wants to find the time-dependent propagator corresponding to $q({\bf r}_0,{\bf r},\beta)$; one could call this unknown function $g({\bf r}_0,{\bf r},\beta,t)$. A formalism for finding $g({\bf r}_0,{\bf r},\beta,t)$ would be extremely useful; it would be the field-theoretic analogue of dynamic path integral simulation methods such as ring polymer molecular dynamics \cite{Habershon2013} and centroid molecular dynamics \cite{Voth1996,Roy1999a,Roy1999b}. Other examples of representing quantum particle dynamics by classical polymer dynamics can be found in the references of Andersen \cite{Andersen2019}. It is not the objective of this work to present a formalism for such calculations, that is, a method for finding $g({\bf r}_0,{\bf r},\beta,t)$ will not be provided. Rather, the purpose here is to prove that it is \emph{possible} to represent quantum dynamics in a field-theoretic way by a classical polymer model in five dimensions. This is in contrast to the simulation methods just mentioned which generally treat the extra thermal dimension as a mathematical tool without physical reality. 

\begin{figure}
\includegraphics[width=1.0\textwidth]{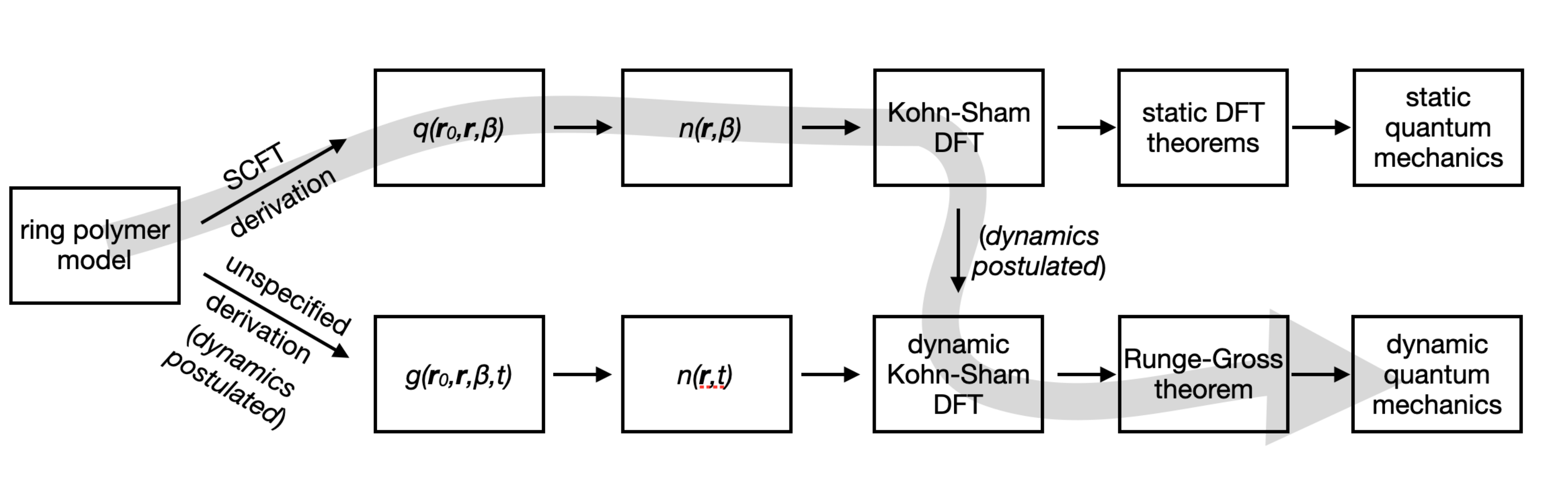}
\caption{A flowchart showing two different methods of connecting the ring polymer picture of quantum particles with time-dependent quantum mechanics. The method used in this letter follows the grey arrow.}
\label{fig:flowchart}
\end{figure}

An alternative to deriving an expression for $g({\bf r}_0,{\bf r},\beta,t)$ is to postulate dynamics directly in terms of a time dependent density $n({\bf r},t)$, as illustrated in figure \ref{fig:flowchart}. 
One can postulate the dynamic quantum particle density expression as the obvious generalization of equation (\ref{n1}):
\begin{equation}
n({\bf r},t) = \frac{1}{V}\sum_{i=1}^\infty f_i \left| \phi_i ({\bf r},t)\right|^2  \label{n2}
\end{equation}
where $f_i$ is the occupancy at an initial time. Postulate (\ref{n2}) is incomplete without a definition for $\phi_i ({\bf r},t)$, which is given by the time-dependent Kohn-Sham equation
\begin{equation}
i\hbar \frac{\partial}{\partial t} \phi_i ({\bf r},t) = \mathcal{H} \phi_i({\bf r},t)   \label{KS2}
\end{equation}
where $\mathcal{H}$ now has a time-dependent field
\begin{equation}
\mathcal{H} = \frac{\hbar^2}{2m} \nabla^2  - w({\bf r},t)  .  \label{H2}
\end{equation}
This time-dependent field would be specified according to the system under study. Equations (\ref{n2})-(\ref{H2}) together form a complete mathematical statement of a single dynamical postulate.\footnote{One could include spin explicitly in the equations, or use time-dependent current density functional theory, but these features would unnecessarily complicate the presentation.} This postulate conserves all quantities such as mass, particle number, momentum, etc., and the equations are of the same form as Kohn-Sham TDDFT \cite{Gross2006,Li1985a,Li1985b}.\footnote{It is assumed that the temperature is fixed throughout the dynamics, but that can be relaxed following \cite{Burke2016}.} Just like in TDDFT, the exchange-correlation potential in $w({\bf r},t)$ will depend, in general, on the history of the time-dependent density $n({\bf r},t)$ and the initial state and so, in principle, memory effects are included in the formalism. In practice, memory effects are often ignored in TDDFT by using the adiabatic approximation. Since equations (\ref{n2})-(\ref{H2}) are the same as in TDDFT, they will reproduce all predictions of non-relativistic quantum mechanics through the theorem of Runge and Gross \cite{Runge1984} and finite temperature generalizations \cite{Li1985a,Burke2016} but from a five dimensional classical polymer model of quantum particles. This is the main point of this letter. Rather than postulating the dynamical nature of the model and attempting to derive the mathematics, the mathematics have been postulated, and now all that is left is to verify that the postulated dynamics remain compatible with the classical ring polymer interpretation. 

To this end, $\nu$-representability will be assumed for both TDDFT and static DFT throughout this work. The theorems of DFT guarantee a one-to-one mapping between a potential function $\nu$ and a density function $n$. This means that given a Hamiltonian with a potential, there is a unique density that corresponds to it. Likewise, that density is generated \emph{only} by that potential; no other potential (to within a spatially constant quantity) will give that same density function. However, an arbitrarily chosen density function does not necessarily correspond to \emph{any} potential. Such a density is non-$\nu$-representable. Research into the mathematical foundations of DFT, including aspects of $\nu$-representability, is a large and important area --- see references \cite{Gonis2016,Ruggenthaler2015} for example for discussions of $\nu$-representability, ensemble $\nu$-representability, non-interacting $\nu$-representability, $N$-representability and density-potential mapping. These issues can be left beyond the scope of this work since, although a full proof of $\nu$-representability is still an open problem \cite{Ruggenthaler2015}, it can be proven under fairly mild assumptions \cite{Gross2006,vanLeeuwen2006a}. Quoting Ayer and Liu \cite{Ayers2007}, ``At this stage the authors view the problem as solved, albeit perhaps not completely understood.'' From a practical perspective, the assumption of $\nu$-representability for both TDDFT and static DFT is reasonable.

Many actual calculations using (\ref{n2})-(\ref{H2}) exist --- all of Kohn-Sham TDDFT is evidence of the practicality of these expressions \cite{Casida2012}. However none of these results can be taken as evidence for a 5D classical polymer model of quantum particles since TDDFT is typically derived through other means \cite{vanLeeuwen2006b}. The dynamic postulate (\ref{n2})-(\ref{H2}) can be viewed as a phenomenological expression of the dynamics, and one would like to confirm that it is consistent with the microscopic polymer picture rigorously derived for the static case. The postulate (\ref{n2})-(\ref{H2}) is consistent in the static limit with the classical model of thermal threads since (\ref{n2})-(\ref{H2}) are constructed to reduce correctly to the static case. This is a necessary, but not sufficient condition for compatibility. It is also required to establish that stationary, equilibrium points, of a dynamical trajectory are not special cases, with the polymer model applying only to those points. 

The polymeric nature of the static model is captured by the modified diffusion equation (\ref{diff3}) which statistically expresses the polymer degrees of freedom through the real-valued propagator $q({\bf r}_0,{\bf r},\beta)$. The static density is proportional to this quantity through equation (\ref{n3}), that is, $n({\bf r},\beta) \propto q({\bf r},{\bf r},\beta)$. For the polymer model to be dynamically consistent with equations (\ref{n2})-(\ref{H2}), there should exist a real-valued time-dependent propagator $g({\bf r}_0,{\bf r},\beta,t)$ that is proportional to the time-dependent density, $n({\bf r},t) \propto g({\bf r},{\bf r},t)$. Such a propagator is known to exist for the set of equations (\ref{n2})-(\ref{H2}).\footnote{See equations 3.12 and 3.13 of reference \cite{vanLeeuwen2006b}} Therefore the 5D classical ring polymer model of quantum particles is compatible with the postulated dynamics (\ref{n2})-(\ref{H2}), which are in turn identical to the equations of TDDFT. So, through the Runge-Gross theorem, all predictions of dynamic quantum mechanics are compatible with a 5D classical ring polymer model arising from only three postulates.

The three postulates given here also function to eliminate other assumptions that are external to quantum physics. The postulates require that there are five dimensions: three spatial, one temporal and one thermal. In the classical limit however, it has been proven that the ring polymer equations reduce to \emph{classical} DFT. See appendix C of reference \cite{Thompson2019}. In other words, particles become point-like, shrinking to lose their contour aspects, and no longer depend on the fifth thermal dimension. In the classical limit, the thermal dimension still exists, but no quantities depend on it any longer. This is called the \emph{cylinder condition}. A five dimensional classical theory subject to a cylinder condition derived following the methods of general relativity was published by Theodor Kaluza a century ago \cite{Kaluza1921}. In what is known as the ``Kaluza miracle'', this theory spontaneously produced both Maxwell's equations of electromagnetism and 4D general relativity from 5D vacuum relativity equations \cite{Wesson1997}. It has been criticized for not identifying the fifth dimension nor justifying the cylinder condition \cite{Wesson1997}. As discussed in this letter, the ring polymer interpretation of quantum foundations is five dimensional, with the fifth dimension clearly identified, and in the classical limit the cylinder condition automatically emerges. It is thus a ``non-compactified'' theory in which the cylinder condition is approximately true in the classical limit \cite{Wesson1997}. 

Other authors have suggested treating quantum particles as extended objects using completely different mathematics \cite{Andersen2019,Raju1981}. In particular, the approach of Andersen \cite{Andersen2019} uses quantum field theory rather than DFT, but is still based on a filament that exists in a physically real fifth dimension arising from Feynman's path integral expression of the partition function. Andersen is able to develop a perturbation theory in Feynman diagrams and show the development of loops and vacuum contributions once they are integrated over the fifth dimension. However alternative postulates to quantum mechanics are not given and the cylinder condition does not spontaneously emerge as it does in the DFT approach, so the relation with electrodynamics through the Kaluza idea is not as immediate.

The connection with classical electrodynamics and a geometric picture of physics highlights a limitation in the DFT approach -- there is a disconnect with the standard model which is based on quantum field theory. This could perhaps be partially alleviated through relativistic DFT, for which there is another existence theorem, analogous to the theorems of non-relativistic DFT, relating the one-particle density to quantum electrodynamics \cite{Engel2017}. One could therefore consider replacing the time-dependent Kohn-Sham equation dynamical postulate with the Dirac-Kohn-Sham equation while maintaining the physical interpretation of quantum particles as threads in 5D. This doesn't help more broadly with the electroweak or strong interactions, nor does the DFT approach help in any way with quantum gravity. Also, while spin can be easily included in calculations within this formalism, a physical interpretation of spin, similar to the extended thread picture of quantum particles, is not attempted, although there is some scope for this following the ideas of Belinfante \cite{Belinfante1939} as described by Ohanian \cite{Ohanian1985}. These issues are all beyond the scope of this letter and are mentioned only to underline some limitations of the ring polymer interpretation which is confined solely to non-relativistic quantum mechanics. 

In summary, a classical ring polymer interpretation of quantum physics can connect the classical limit of quantum mechanics directly to gravity and electromagnetism through geometry, and remove the need for postulating Maxwell's equations separately. This is in addition to reducing the number of postulates required for quantum mechanics to three, and eliminating the measurement problem. Since the whole theory fits neatly within classical statistical mechanics with an extra thermal dimension, there is improved economy of thought, and improved explanatory power. This recommends the 5D classical statistical mechanical ring polymer theory for consideration as an interpretation of quantum mechanics.

\section*{Acknowledgements}
The author acknowledges helpful discussions with R. A. Wickham and P. Le Maitre. This research was financially supported by the Natural Sciences and Engineering Research Council of Canada (NSERC).

\bibliography{DFTbibliography7}

\end{document}